\documentclass[aps,pra,twocolumn,10pt,superscriptaddress,longbibliography,floatfix]{revtex4-1}
\usepackage{bm}
\usepackage{hyperref}
\usepackage{amsmath}
\usepackage{amssymb}
\usepackage[dvipsnames]{xcolor}
\usepackage{ulem}
\usepackage{xcolor}
\definecolor{respblue}{RGB}{0,70,180}
\definecolor{newblue}{RGB}{0,70,180}

\usepackage{amsmath,amssymb,amsfonts}

\usepackage{braket}

\providecommand{\ket}[1]{\ensuremath{\left| #1 \right\rangle}}

\usepackage{amsmath,amssymb,amsfonts}

\usepackage{braket}

\begin{document}

\title{\textbf{{Statistical Mechanics from Quantum Envariance and Exchange Symmetry}} 
}% 

\author{Amul Ojha}
\thanks{amul21@iiserb.ac.in}
\affiliation{Department of Physical Sciences, Indian Institute of Science Education and Research (IISER) Bhopal, Madhya Pradesh 462066, India}

\author{Shubhit Sardana}\thanks{shubhit21@iiserb.ac.in}
\affiliation{Department of Physical Sciences, Indian Institute of Science Education and Research (IISER) Bhopal, Madhya Pradesh 462066, India}

\author{Arnab Ghosh}
\thanks{arnab@iitk.ac.in}
\affiliation{Indian Institute of Technology (IIT) Kanpur, Uttar Pradesh 208016, India}
\date{\today}

\begin{abstract}
We build on the foundational work of Deffner and Zurek
[S. Deffner and W. H. Zurek, \href{https://doi.org/10.1088/1367-2630/18/6/063013}{New J. Phys. \textbf{18}, 063013 (2016)}]
to show how central equilibrium structures of statistical mechanics can be understood within standard quantum mechanics using the concept of envariance (environment-assisted invariance).
In particular, we show how the Binomial, Poisson, and Gaussian distributions naturally emerge from entangled system--environment states.
We revisit the Gibbs paradox from a quantum-information perspective, demonstrating that the standard Sackur--Tetrode entropy and its $1/N!$ factor arise from indistinguishability enforced through entanglement with an environment, without introducing additional thermodynamic corrections.
Within the same framework, we analyze ionization equilibrium and show how the classical Saha equation is recovered, while clarifying how indistinguishability enters through an entanglement-induced reduction of permutation redundancy. Assuming the standard exchange symmetries of identical quantum particles, we further show how the Bose--Einstein and Fermi--Dirac distributions follow as the equilibrium weighting of symmetry-allowed occupation configurations.
Overall, our results support the view that equilibrium statistical mechanics can be consistently interpreted as an emergent consequence of quantum information–theoretic structure and symmetry, rather than as a collection of independent phenomenological postulates.
\end{abstract}

\maketitle
\section{\label{sec:level1}Introduction}

Statistical mechanics can be reformulated from the perspective of quantum information, as
outlined in the seminal work of Deffner and Zurek
\cite{deffner2016,brukner1999information,popescu2006,zurek2009quantum,sone2025no_go}.
Historically, however, the subject was established by Maxwell
\cite{Maxwell1871TheoryOfHeat,Maxwell1890ScientificPapers}, Boltzmann
\cite{boltzmann1896,zurek2009quantum} and Gibbs
\cite{gibbs1902,vinjanampathy2016quantum}, who formulated the statistical description of
many-particle systems based on probabilistic principles. In particular, the postulate of
equal a priori probabilities for an isolated system in equilibrium has long served as a
cornerstone of the theory. Although remarkably successful, these principles are typically
introduced as assumptions, motivated more by their explanatory power than by derivation
from the first principles \cite{schrodinger1952}.This has left open the question of whether statistical mechanics admits a deeper foundation
within fundamental physics.

Recent developments suggest that this foundation may be rooted in the principles of quantum
mechanics. The concept of envariance, which was first introduced by Zurek
\cite{zurek2003} and later expanded by Deffner and Zurek
\cite{deffner2016}, demonstrates that objective probabilities and equilibrium structures
emerge naturally from the symmetries of quantum entanglement \cite{brandao2013}.
A detailed review on this topic can be found in Ref.~\cite{sebastian2019book}.
Within this framework, envariance provides a symmetry-based justification for assigning
equal probabilities to all configurations in the microcanonical ensemble, without appeal
to subjective ignorance.

In what follows, we have provided a unified derivation of the binomial, Poisson, and Gaussian distributions from a single quantum-mechanical model, highlighting the predictive efficacy and internal consistency of this approach. It shows that the statistics that govern discrete events (like particle
counts) and continuous variables are not separate things, but are two sides of the same
quantum reality. From this perspective, classical probability theory appears as an effective description
emerging from more fundamental quantum information-theoretic constraints.

Our investigation proceeds as follows:
\begin{enumerate}
    \item \textbf{Principle of envariance} --- In Sec.~\ref{Sec-2}, we review the concept of envariance, following Zurek~\cite{zurek2003} and Deffner (Foundations of statistical mechanics from symmetries of entanglement~\cite{deffner2016,deffner2016}), and explain its role in justifying the equal a priori probability postulate of the microcanonical ensemble.

    \item \textbf{Probability distributions} --- We then derive the Binomial, Poisson, and
    Gaussian distributions in Sec.~\ref{Sec-3} as natural consequences of quantum
    measurement statistics, showing that classical randomness is unnecessary; instead,
    statistical structure arises from entangled quantum states.

    \item \textbf{Resolution of the Gibbs paradox} --- Following the work of Deffner
    and Zurek \cite{deffner2016,sebastian2019book}, we then briefly review the canonical
    ensemble and Boltzmann distribution in Sec.~\ref{Sec-4}, as the natural statistical
    description of a quantum system coupled to a heat bath. This review builds on the microcanonical
    framework and connects seamlessly to thermodynamics, making the presentation
    self-contained. Next, we revisit the Gibbs paradox
    \cite{gibbs1902,sackur1911,tetrode1912} in Sec.~\ref{Sec-5}, showing that the spurious
    entropy of mixing for identical particles is canceled by an entanglement-based
    correction enforcing indistinguishability. We illustrate how this correction is
    negligible for gases under ordinary conditions but significant in the ultracold regime.

   \item \textbf{Quantum statistics} --- In Sec.~\ref{Sec.-6}, we clarify how Bose--Einstein and Fermi--Dirac statistics arise from the combination of exchange symmetry and envariance-based equilibrium weighting. This provides the necessary foundation for treating indistinguishable particles in various regimes.

   \item \textbf{Modified Saha equation} --- Finally, we apply this framework to ionization equilibrium in Sec.~\ref{Sec.-7}. We derive a quantum correction to the classical Saha equation \cite{saha1921(2)}, demonstrating how envariance clarifies the role of indistinguishability in chemical potential and ionization rates.
\end{enumerate}

Finally, this paper aims to present that the laws and distributions of statistical mechanics
are not separate ad hoc constructs, but are interconnected results that emerge from the
symmetries and information-theoretic properties of the quantum world
\cite{yuan2010,bera2019}.With this perspective, we seek to clarify the quantum-mechanical foundations of statistical
mechanics without any extra assumptions.The principle of envariance, as subsequently refined in later works
\cite{zurek2014quantumorigin,zurek2018einselection,vedral2012classicalworld,
zurek2022envarianceupdate,barnum2000information}, continues to provide a bridge between
quantum foundations and statistical inference.

\section{Review the principle of Envariance~\cite{deffner2016}}\label{Sec-2}

Consider a quantum system $S$ that is \textit{maximally entangled} with an environment
$E$.
This assumption corresponds to an idealized equilibrium limit in which the system
is fully correlated with a sufficiently large environment, and it provides the symmetry setting
required for the envariance arguments discussed below.
Let $|\Psi_{SE}\rangle$ denote the composite state of the system and environment in the
Hilbert space $\mathcal{H}_S \otimes \mathcal{H}_E$.
The state $|\Psi_{SE}\rangle$ is said to be \textit{envariant} under a unitary operation
$U_S$ acting on the system $S$ \textit{iff} there
exists another unitary operation $U_E$ acting on the environment $E$ that can
restore the original state \cite{zurek2003}.
Mathematically:
\begin{equation}
    (U_S \otimes I_E)|\Psi_{SE}\rangle = |\Phi_{SE}\rangle
    \label{1}
\end{equation}
\begin{equation}
    (I_S \otimes U_E)|\Phi_{SE}\rangle = |\Psi_{SE}\rangle
     \label{2}
\end{equation}
The key insight is that the action of $U_S$ on the system can be perfectly compensated
by an action $U_E$ on the environment alone,
reflecting a symmetry property of the global entangled state rather than a
dynamical relaxation process.

\subsection{Example: Two-Level System}

Let system $S$ and environment $E$ be two-level systems (qubits).
The eigenstates for $S$ are $\{|\uparrow\rangle_S, |\downarrow\rangle_S\}$ and for $E$
are $\{|\uparrow\rangle_E, |\downarrow\rangle_E\}$.
Assume that they are prepared in a maximally entangled Bell state~\cite{sebastian2019book}:
\begin{equation}
    |\Psi_{SE}\rangle = \frac{1}{\sqrt{2}} \left(
    |\uparrow\rangle_S \otimes |\uparrow\rangle_E +
    |\downarrow\rangle_S \otimes |\downarrow\rangle_E
    \right)
     \label{3}
\end{equation}
Let us consider a `SWAP' (or spin-flip) operation $U_S$ on the system $S$,
which corresponds to the $\sigma_x$ Pauli operator, such that
\begin{equation}
    \begin{split}
    (U_S \otimes I_E) |\Psi_{SE}\rangle
    &= \frac{1}{\sqrt{2}} \Big(
    |\downarrow\rangle_S \otimes |\uparrow\rangle_E \\
    &\quad + |\uparrow\rangle_S \otimes |\downarrow\rangle_E
    \Big)
    = |\Phi_{SE}\rangle
     \label{4}
    \end{split}
\end{equation}
The state has changed.
Now, can this be undone by acting only on the environment?
Let us apply a corresponding `SWAP' operation $U_E$ on the environment $E$:
\begin{equation}
    \begin{split}
    (I_S \otimes U_E) |\Phi_{SE}\rangle
    &= \frac{1}{\sqrt{2}} \Big(
    (I_S \otimes U_E) |\downarrow\rangle_S \otimes |\uparrow\rangle_E \\
    &\quad + (I_S \otimes U_E) |\uparrow\rangle_S \otimes |\downarrow\rangle_E
    \Big)\\
    & = \frac{1}{\sqrt{2}} \left(
    |\downarrow\rangle_S \otimes |\downarrow\rangle_E +
    |\uparrow\rangle_S \otimes |\uparrow\rangle_E
    \right)\\
    & = |\Psi_{SE}\rangle
     \label{5}
    \end{split}
\end{equation}
The swap on $E$ has restored the original \textit{global state} without touching $S$.
Therefore, the state $|\Psi_{SE}\rangle$ is envariant under the spin-flip operation $U_S$
\cite{deffner2016,zurek2003,sebastian2019book}.

\paragraph*{Implication for Probabilities.}
The fact that the states $|\uparrow\rangle_S$ and $|\downarrow\rangle_S$ can be swapped
by a transformation that leaves the global state invariant implies that any measurement
on the system must assign equal probability to these two outcomes
within the maximally entangled equilibrium scenario considered here.
The local probabilities of the swapped states are exchanged by $U_S$ but restored by
$U_E$, leaving all observable properties of the reduced state invariant; hence the
probabilities associated with these outcomes must be equal.
This provides a symmetry-based justification for the principle of \textit{equal a priori}
probabilities in the microcanonical ensemble
\cite{deffner2016,boltzmann1896,gibbs1902}.
Recent developments in decoherence and quantum Darwinism
\cite{zurek2018quantumdarwinismreview,korbicz2021quantumdarwinism,
busch2016quantumreality,zurek2023quantumdarwinism}
further support this interpretation of statistical mechanics as a theory of emergent
classicality.

Before proceeding further with our reconstruction of statistical mechanics based on envariance, it is useful to briefly emphasize the central idea following in the footsteps of Zurek, Popescu and Deffner~\cite{zurek2003,popescu2006,deffner2016}. Traditional derivations of the microcanonical, canonical, and grand canonical equilibrium have typically invoked concepts such as probability, ergodicity, ensembles, randomness, and the principle of indifference. Yet, within the framework of statistical physics, these notions are not always defined with complete mathematical rigor.. Historically, pioneers such as Maxwell and Boltzmann grappled with these conceptual challenges~\cite{Maxwell1871TheoryOfHeat,Maxwell1890ScientificPapers,boltzmann1896}. The modern formal structure of statistical mechanics was primarily due to Gibbs, who adopted a probabilistic framework without directly resolving these foundational issues~\cite{gibbs1902}.

By contrast, present approach relies solely on a quantum symmetry arising from entanglement --- namely, envariance. In this way, we provide a reformulation of the fundamental principles of statistical mechanics within a purely quantum-mechanical framework.

\subsection{Microcanonical equilibrium and probability distributions from envariance}

\paragraph{Envariance and the Schmidt decomposition.}
Consider a pure state of a composite system $S+E$ written in its Schmidt form
\begin{equation}
    {\psi_{SE}} = \sum_{k} a_k |{S_k}\rangle \otimes |{E_k}\rangle,
    \label{6}
\end{equation} 
where $\{|S_k\rangle\}$ and $\{|E_k\rangle\}$ are orthonormal bases for the system $S$
and environment $E$, respectively; $a_k\in\mathbb{C}$ are the Schmidt coefficients
(which can be chosen real and non-negative by an appropriate phase convention);
the sum runs over the Schmidt rank.
The Schmidt decomposition provides a canonical bipartite representation, and all
local statistical properties of $S$ are fully determined by $\{a_k\}$ and
$\{|S_k\rangle\}$~\cite{deffner2016,sebastian2019book}.

\paragraph{Maximal envariance $\Rightarrow$ equal Schmidt weights.}
A state is said to be \emph{maximally envariant} if for every unitary transformation
$U_S$ acting on the system $S$ there exists a corresponding unitary $U_E$ acting on
the environment $E$ such that Eqs.~\eqref{1}–\eqref{2} are satisfied.
This condition characterizes an idealized equilibrium situation in which
the system is fully entangled with a sufficiently large environment, ensuring maximal
symmetry under local transformations of $S$.
If Eqs.~\eqref{1}–\eqref{2} hold for all $U_S$ acting within the support of $\rho_S$,
then all nonzero Schmidt coefficients must have the same modulus,
\begin{equation}
    |a_k| = \frac{1}{\sqrt{\Omega}}\quad \forall\; k \ \text{(within the active subspace),}
    \label{7}
\end{equation}
where $\Omega$ is the number of Schmidt terms, corresponding to the effective
dimension of the degenerate subspace.
For the $N$-qubit example discussed below, $\Omega=2^N$ equals the dimension of the
support of $\rho_S$ when the entire computational basis participates.
Equal Schmidt magnitudes are both necessary and sufficient for the existence of
compensating environment unitaries $U_E$ that undo arbitrary local unitaries $U_S$
acting within this support, which justifies the term ``maximal envariance.''

\paragraph{Equiprobability from envariance.}
Tracing out the environment yields the reduced state of the system,
\begin{equation}
    \rho_S = \mathrm{Tr}_E \big( |{\psi_{SE}}\rangle\langle{\psi_{SE}|} \big)
    = \sum_k |a_k|^2 |{S_k}\rangle\langle{S_k}|,
    \label{eq:reduced}
\end{equation}
which is diagonal in the Schmidt basis.
If the global state is maximally envariant, then $|a_k|^2 = 1/\Omega$ according to
Eq.~\eqref{7}, and therefore
\begin{equation}
    \rho_S = \frac{1}{\Omega}\sum_{k=1}^{\Omega} |{S_k}\rangle\langle{S_k}|,
    \qquad
    P(S_k)=\frac{1}{\Omega}.
    \label{9}
\end{equation}
Here $P(S_k)=\langle S_k|\rho_S|S_k\rangle$ denotes the Born probability
\cite{Born1926ZPhys,WheelerZurek1983QuantumMeasurement}
for obtaining the outcome associated with $|S_k\rangle$.
Because local permutations or swaps among the $|S_k\rangle$ states can be exactly
compensated by corresponding transformations acting on $E$, no outcome can be
distinguished by operations confined to the system alone.
Consequently, all outcomes occur with equal probability
within the maximally envariant equilibrium regime considered here.

\paragraph{Physically motivated Hamiltonian.}
To provide physical intuition for how such maximally envariant states may arise,
consider a symmetric, pairwise system–environment interaction of the form
\begin{equation}
    H_{SE} = \sum_{i=1}^N \left(
    H_{S_i}\otimes I_{E_i} + I_{S_i}\otimes H_{E_i}
    + g\, A_i\otimes B_i
    \right),
    \label{eq:HSE}
\end{equation}
where $S=\bigotimes_i S_i$ and $E=\bigotimes_i E_i$;
$H_{S_i}$ and $H_{E_i}$ are local Hamiltonians; and $A_i\otimes B_i$ denotes an
interaction operator acting identically on each system–environment pair
$(S_i,E_i)$ with coupling strength $g$.
Such symmetric interactions favor permutation-symmetric entanglement structures and,
when acting within a degenerate energy subspace, can drive the composite system
toward states of the form in Eq.~\eqref{6} with approximately equal Schmidt weights.
We emphasize that this construction is intended as a physically motivated
illustration rather than a general proof of equilibration under arbitrary dynamics.
The underlying envariance-based symmetry arguments have also been explored within
information-theoretic approaches and quantum resource theories
\cite{jaynes1957information, brukner1999information, cabello2012foundations,
ferrari2025, minagawa2025, d_alessio2015, picatoste2025, popovic2023}.

\paragraph{Logical status and scope of envariance.}
It is important to clarify that envariance does not replace the axioms of quantum
mechanics, but rather exploits their unitary symmetry structure.
Consider a system--environment state written in Schmidt form as
$|\Psi_{SE}\rangle = \sum_k a_k \, |S_k\rangle \otimes |E_k\rangle$.
Maximal envariance under all system permutations implies that for any swap
$|S_k\rangle \leftrightarrow |S_{k'}\rangle$ there exists a compensating unitary acting
only on the environment, which is possible if and only if the corresponding Schmidt
coefficients satisfy $|a_k| = |a_{k'}|$.
As a consequence, all nonzero Schmidt coefficients must have equal modulus,
$|a_k|^2 = 1/\Omega$.
Tracing out the environment then yields a reduced system state of the form
$\rho_S = \Omega^{-1}\sum_k |S_k\rangle\langle S_k|$, so that the probabilities
$P(S_k)=\langle S_k|\rho_S|S_k\rangle$ are equal.
In this restricted equilibrium (maximally envariant) setting, equiprobability therefore
follows as a consequence of quantum entanglement and symmetry, rather than as an
independent postulate.

The assumption of maximal envariance corresponds to an idealized equilibrium limit,
analogous to the use of infinite heat baths or the thermodynamic limit in standard
statistical mechanics.
Mathematically, maximal envariance characterizes fixed points of the reduced dynamics:
a reduced state proportional to the identity on its support is invariant under all
unitaries acting within that support and therefore admits no locally distinguishable
structure.
Deviations from maximal envariance may be parameterized by Schmidt weights of the form
$|a_k|^2 = \Omega^{-1} + \delta_k$ with $\sum_k \delta_k = 0$, in which case the trace
distance between the reduced state and the maximally mixed state is bounded by a quantity
of order $\max_k |\delta_k|$.
Thus, approximate envariance implies approximate equiprobability, establishing the
robustness of the argument under weak deviations from ideal entanglement.
Non-equilibrium or weakly entangled regimes fall outside the scope of the present
derivation, just as they do for conventional equilibrium ensembles; envariance is
intended to characterize equilibrium structure rather than dynamical relaxation.

\paragraph{Consistency of environmental assumptions.}
Several derivations in this work employ an idealized environment that is
(i) maximally entangled with the system,
(ii) dynamically passive at the level of reduced system observables,
and (iii) capable of recording permutation or occupation information.
We clarify here that these assumptions are mutually consistent within a well-defined
equilibrium limit and are not stronger than those routinely employed in standard
statistical mechanics.

First, \emph{dynamical passivity} does not imply the absence of interaction, but rather
the invariance of the reduced system state under all local unitary transformations.
If the reduced state satisfies $\rho_S = \Omega^{-1} I_{\mathrm{supp}}$, then for any
unitary operator $U_S$ acting within the support one has
$U_S \rho_S U_S^\dagger = \rho_S$, implying that local observables cannot detect the
action of such transformations.
Thus, the system appears dynamically passive at the level of reduced observables, even
though entangling interactions with the environment are present at the global level.

Second, the ability of the environment to record permutation or occupation information
requires only that distinct system configurations be correlated with approximately
orthogonal environmental states.
For an environment with Hilbert-space dimension much larger than that of the system,
typical overlaps between distinct environmental states scale inversely with the square
root of the environment dimension, so that orthogonality emerges rapidly as the bath
size increases.
This is precisely the mechanism underlying decoherence and the formation of robust
classical records in quantum Darwinism \cite{zurek2003decoherence, zurek2009,korbicz2021quantumdarwinism}.

Finally, maximal envariance corresponds to an equilibrium idealization analogous to the
microcanonical ensemble or the thermodynamic limit.
Small deviations from perfect entanglement lead to reduced states whose deviation from
the maximally mixed form is controlled by the same Schmidt-weight deviations
$\delta_k$, implying that equiprobability remains approximately valid.
Thus, approximate entanglement implies approximate envariance and hence approximate
equiprobability.

We emphasize that the envariance framework is not intended to describe generic
non-equilibrium dynamics or weakly entangled transient regimes.
Rather, it characterizes the structural properties of equilibrium states, in the same
sense that conventional ensembles do not describe the microscopic path to
thermalization but only its asymptotic form.

\section{Probability distributions from quantum envariance principle}
\label{Sec-3}

\subsection{Binomial Distribution}
\subsubsection{Maximally entangled case}

First, we consider $N$ identical two-level subsystems (qubits) and label each system basis
by $|{\uparrow}\rangle$ and $|{\downarrow}\rangle$.
The maximally envariant state over the full computational subspace can be written as
\begin{equation}
    |{\psi_{SE}\rangle} = \frac{1}{\sqrt{\Omega}} \sum_{\mathbf{S}}
    |{S_1 S_2 \cdots S_N}\rangle \otimes |{E_{S_1\cdots S_N}}\rangle,
    \label{11}
\end{equation}
where the sum runs over all binary strings $\mathbf{S}=(S_1,\dots,S_N)$ and
$\Omega=2^N$.
The states $|{E_{S_1\cdots S_N}}\rangle$ form an orthonormal basis of the environment
that records which system configuration occurs.
Equation~\eqref{11} is precisely the equal-weight Schmidt decomposition across the
$2^N$ computational states,
corresponding to a maximally envariant equilibrium configuration as
defined in Sec.~\ref{Sec-2}.

We now ask: what is the probability $P(n)$ of finding exactly $n$ subsystems in the
state $|{\uparrow}\rangle$?
The number of computational basis strings with Hamming weight $n$
(i.e., exactly $n$ up-spins) is $\binom{N}{n}$.
Since envariance enforces equiprobability of all $\Omega$ fine-grained configurations
[cf.~Eq.~\eqref{9}], the probability of the coarse-grained outcome $n$ is
\begin{equation}
    P(n) = \frac{\binom{N}{n}}{\Omega}
         = \frac{\binom{N}{n}}{2^N}.
    \label{12}
\end{equation}
Here $N$ denotes the total number of qubits and $n$ the number of
$|{\uparrow}\rangle$ outcomes.
Thus, the binomial distribution emerges directly from the combination of
envariance-induced equiprobability at the fine-grained level and combinatorial
counting of coarse-grained events.

\subsubsection{General binomial distribution via ancilla embedding}

To obtain a biased single-qubit probability $p\neq 1/2$ while remaining within an
envariance-based framework, we embed the physical system into a larger fine-grained
Hilbert space by introducing an auxiliary (ancilla) system.
Choose positive integers $m$ and $M$ with $0<m<M$ and define
\begin{equation}
    p = \frac{m}{M}.
    \label{eq:p_fraction}
\end{equation}
We construct a uniform superposition over $M^N$ fine-grained states such that each
physical $|{\uparrow}\rangle$ state is correlated with $m$ orthogonal ancilla labels,
while each $|{\downarrow}\rangle$ state is correlated with $(M-m)$ orthogonal labels.
The number of fine-grained states corresponding to exactly $n$ physical up-spins is
\begin{equation}
    \#(\text{fine states with }n)
    = \binom{N}{n}\, m^n (M-m)^{N-n}.
    \label{eq:fine_count}
\end{equation}
Dividing by the total number $M^N$ of equiprobable fine-grained states yields
\begin{equation}
    P(n)
    = \frac{\binom{N}{n}\, m^n (M-m)^{N-n}}{M^N}
    = \binom{N}{n}\, p^n (1-p)^{N-n}.
    \label{15}
\end{equation}
The ancilla embedding maps unequal coarse-grained probabilities onto equal-weight
fine-grained microstates, allowing envariance to enforce equiprobability at the
expanded level.
This construction serves as a technical device for representing biased
probabilities without introducing additional physical assumptions beyond the
enlarged Hilbert space.

\paragraph*{Remarks.}
In all cases considered above, the probabilistic structure arises from envariance
applied to maximally entangled system--environment states.
Maximally envariant (equal-Schmidt) states enforce equiprobability, while auxiliary
embeddings allow this argument to be extended to generic biased probabilities.
No appeal is made to subjective ignorance; probabilities arise from
symmetry properties of entangled quantum states.

\subsection{The Poisson Limit}

The Poisson distribution arises as a limiting case of the binomial distribution.
In the limit $N\to\infty$, $p\to0$, with $\lambda=Np$ held fixed, the binomial mass
function becomes
\begin{equation}
    P(n)
    = \lim_{\substack{N\to\infty\\ p\to0}}
      \binom{N}{n}p^n(1-p)^{N-n}
    = \frac{\lambda^n e^{-\lambda}}{n!}.
    \label{16}
\end{equation}
This limit describes rare but finite events occurring across many trials, a scenario
common in quantum measurement of low-probability processes.
In quantum optics, the Poisson distribution characterizes the photon statistics of
coherent states~\cite{glauber1963,cohen1991}.
Its appearance here reflects the large-$N$ limit of the same envariance-based
binomial structure.
A detailed derivation is provided in Appendix~A.

\subsection{Gaussian Approximation of Measurement Outcomes}

We aim to demonstrate that the probability distribution of measurement outcomes on a system of a large number of identical qubits approaches a Gaussian distribution. This is a quantum mechanical analogue of the de Moivre-Laplace theorem ~\cite{nielsen2000,breuer2002}, which states that the binomial distribution can be approximated by a normal (Gaussian) distribution under certain conditions.

Consider a system of $N$ identical qubits, each independently prepared in state $|\psi_{SE}\rangle = \sqrt{p}|\uparrow\rangle + \sqrt{1-p}|\downarrow\rangle$. The global state is a product of these states~\cite{zurek2003}:
\begin{equation}
    |\psi_{SE}\rangle = \bigotimes_{i=1}^{N} \left( \sqrt{p}|\ket{\uparrow}\rangle_i + \sqrt{1-p}|\ket{\downarrow}\rangle_i \right)
\label{17}\tag{17}
\end{equation}
where $p$ is the probability that a single qubit being in the state $|\uparrow\rangle$ and $1-p$ is the probability that a single qubit being in the state $|\downarrow\rangle$. 

Expanding the tensor product yields a sum over all computational basis strings. Every basis string with Hamming weight $n$ appears with amplitude $p^{n/2}(1-p)^{(N-n)/2}$. As there are $\binom{N}{n}$ such strings, collecting terms by $n$ gives
\begin{equation}
|\psi_{SE}\rangle
= \sum_{n=0}^N \left[p^{n/2}(1-p)^{(N-n)/2}
\sum_{\substack{\text{strings}\\ \text{of weight } n}} |\{x_1\ldots x_N\}\rangle\right]  
\tag{18}\end{equation}
Recognizing the symmetric sum as $\sqrt{\binom{N}{n}}\,|{D_n^{(N)}}\rangle$ yields~\cite{nielsen2000}
\begin{equation}
    |\psi_{SE}\rangle = \sum_{n=0}^{N} \sqrt{\binom{N}{n}} p^{n/2} (1-p)^{{(N-n)}/2}| \,D_n^{(N)}\rangle,\tag{19}
\label{30}\end{equation}
where $|D_n^{(N)}\rangle$ is the normalized Dicke state with Hamming weight (number of $|\uparrow\rangle$ outcomes) $n$~\cite{nielsen2000}. The normalized symmetric Dicke state of $N$ qubits with Hamming weight $n$ is
\begin{equation}
|D_n^{(N)}\rangle = \frac{1}{\sqrt{\binom{N}{n}}}
\sum_{\substack{\text{weight-}n\\ \text{bit strings}}} \big( |x_1\rangle \ |x_2\rangle \ \cdots |x_N\rangle \big)
\label{20}\tag{20}
\end{equation}
where the sum runs over all the distinct permutations of the bitstring containing $n$ copies of $|\uparrow\rangle$ and $N-n$ copies of $|\downarrow\rangle$. The prefactor ensures $\langle D_n^{(N)} | D_n^{(N)} \rangle = 1$. In this expression, each $|x_1\rangle |x_2\rangle \dots |x_N\rangle$ is the basis vector, and each $x_i$ can be $|\uparrow\rangle$ or $|\downarrow\rangle$. Here `strings of weight $n$' refers to all basis states that contain exactly $n$ spins in the $|\uparrow\rangle$ state and $N-n$ spins in the $|\downarrow\rangle$ state.

Squaring the amplitudes gives the binomial probability law, Eq.\eqref{12}. Hence, the probability of observing $n$ excitations (i.e.,  $n$ qubits in $|\uparrow\rangle$) is binomial as in Eq.~\eqref{15}. For large $N$ (with fixed $0<p<1$), the de Moivre-Laplace theorem (or Stirling expansion) gives the Gaussian distribution (see Appendix B)
\begin{equation}
P(n) \approx \frac{1}{\sqrt{2\pi \sigma^2}}\,
\exp\!\Big[-\frac{(n-\lambda)^2}{2\sigma^2}\Big]
\tag{21}\label{21}
\end{equation}
with mean $\lambda = Np$ and variance $\sigma^2 = N p(1-p)$. This hierarchy of statistical distributions is consistent with the modern understanding of quantum statistical mechanics, wherein equilibrium properties are understood to arise from the dual concepts of \textit{typicality} and \textit{eigenstate thermalization}~\cite{popescu2006, brandao2013, bera2019, eisert2015quantum, yuan2010}.

\section{The Canonical State from Quantum Envariance~\cite{deffner2016}}
\label{Sec-4}

We briefly summarize the canonical state and the Boltzmann distribution following
Refs.~\cite{deffner2016,sebastian2019book} by considering a small system of interest
in contact with a large heat bath,with the aim of clarifying how the canonical ensemble emerges from the
envariance-based microcanonical framework rather than being introduced as an
independent postulate.

\paragraph{Setup.}
Separate a large, isolated system $S$ (which is in a microcanonical state) into two
parts:
\begin{itemize}
    \item $\mathcal{C}$: A smaller subsystem of interest.
    \item $\mathcal{B}$: Its complement, which acts as a large heat bath.
\end{itemize}
The total Hamiltonian of the system $S$ is
$
H_{S} = H_{\mathcal{C}} \otimes I_{\mathcal{B}}
+ I_{\mathcal{C}} \otimes H_{\mathcal{B}}
+ H_{\text{int}} .
$
We assume \textit{ultraweak coupling}, meaning the interaction term $H_{\text{int}}$
is negligible for calculating the total energy
$\mathcal{E} \approx \mathcal{E}_{\mathcal{C}} + \mathcal{E}_{\mathcal{B}}$,
but it is necessary to facilitate energy exchange between $\mathcal{C}$ and
$\mathcal{B}$~\cite{pathria2011}.This assumption corresponds to the standard thermodynamic limit in which the bath
is much larger than the subsystem and remains only weakly perturbed by it.

\paragraph{Counting States.}
We start with the maximally envariant microcanonical state for the total isolated
system $S$.
An energy eigenstate of the composite system can be written as a product state
$
|k\rangle_S = |c_k\rangle_{\mathcal{C}} \otimes |b_k\rangle_{\mathcal{B}},
$
where $|c_k\rangle$ and $|b_k\rangle$ are energy eigenstates of $\mathcal{C}$ and
$\mathcal{B}$, respectively.
If $\mathcal{C}$ is found in a specific energy eigenstate $|c_k\rangle$ with energy
$\epsilon_k$, conservation of total energy implies that the bath must have energy
$\mathcal{E}_{\mathcal{B}} = \mathcal{E} - \epsilon_k$.
The number of accessible microstates is therefore
\begin{equation}
n(\epsilon_k) = n_{\mathcal{B}}(\mathcal{E} - \epsilon_k).
\tag{22}
\end{equation}

Because the total system is assumed to be in a maximally envariant microcanonical
state, all global microstates within the energy shell are equally weighted.
As a result, the probability of observing a given subsystem energy $\epsilon_k$
is proportional to the number of compatible bath states.

The degeneracy of the bath $n_{\mathcal{B}}(\mathcal{E}_{\mathcal{B}})$ is related to
its entropy by $S_{\mathcal{B}} = k_B \ln n_{\mathcal{B}}$.
Expanding around the total energy $\mathcal{E}$ yields
$
S_{\mathcal{B}}(\mathcal{E} - \epsilon_k)
\simeq S_{\mathcal{B}}(\mathcal{E}) - \frac{\epsilon_k}{T},
$
assuming the bath has an effectively infinite heat capacity.
Thus,
$
\ln n_{\mathcal{B}}(\mathcal{E} - \epsilon_k)
\simeq \ln n_{\mathcal{B}}(\mathcal{E}) - \beta \epsilon_k .
$

Therefore,
$
\ln n(\epsilon_k) \simeq -\beta \epsilon_k + \text{const},
$
and the probability becomes
\begin{equation}
P(\epsilon_k) \propto e^{-\beta \epsilon_k}.
\tag{23}
\end{equation}

Within the present framework, the Boltzmann distribution thus emerges from
envariance-enforced equiprobability at the level of the isolated system combined
with the extensivity of the bath entropy, rather than being introduced as an
independent assumption.

\paragraph{Interpretation and scope.}

The canonical ensemble obtained here should be understood as an effective
description of a small subsystem embedded in a much larger envariant whole.
As in conventional statistical mechanics, this construction applies to equilibrium
situations and does not address non-equilibrium dynamics or strongly coupled
regimes, which lie beyond the scope of the present work.

This interpretation is consistent with modern approaches to equilibration and
eigenstate thermalization in closed quantum systems
\cite{gogolin2016equilibration,landi2021nonequilibrium,
miller2019quantumthermo,binder2020thermodynamicuncertainty,
guryanova2016thermoresources,mitra2022thermalizationreview}.

\section{Resolution of the Gibbs Paradox via Quantum Indistinguishability and Entanglement}
\label{Sec-5}

The Gibbs paradox arises in the classical statistical mechanics of ideal gases.
When two identical gases are mixed at the same temperature and pressure, the classical
theory predicts an increase in entropy, known as the ``entropy of mixing''
\cite{gibbs1902, sackur1911, tetrode1912}.
This paradox is resolved by considering the quantum-mechanical indistinguishability of
identical particles within the standard framework of quantum statistical mechanics.

\subsection{The Classical Paradox}

Consider mixing two ideal gases, both at temperature $T$ and occupying volume $V$.
Gas~$1$ has $N_1$ particles and Gas~$2$ has $N_2$ particles.
Using the Sackur--Tetrode equation for entropy~\cite{gibbs1902, sackur1911, tetrode1912},
the initial total entropy for identical gases is
\begin{equation}
\begin{split}
S_{\text{initial}} &= N_1 k_B \left( \ln V + \tfrac{3}{2} \ln T + \sigma \right) \\
&\quad + N_2 k_B \left( \ln V + \tfrac{3}{2} \ln T + \sigma \right),
\end{split}
\tag{24}
\end{equation}
where $\sigma$ is a constant arising from the Sackur--Tetrode expression.
After removing the partition, the combined system of $N = N_1+N_2$ particles occupies
a total volume $2V$.
The final entropy is
\begin{equation}
S_{\text{final}} = (N_1+N_2) k_B \left[ \ln (2V) + \tfrac{3}{2} \ln T + \sigma \right].
\tag{25}
\end{equation}
For identical gases, and taking $N_1 = N_2 = N$ for simplicity, the entropy change is
\begin{equation}
\Delta S = S_{\text{final}} - S_{\text{initial}} = 2N k_B \ln 2.
\tag{26}
\end{equation}
This nonzero entropy of mixing for identical gases constitutes the Gibbs paradox.

The resolution lies in quantum statistical mechanics~\cite{huang1987}.
Identical particles are indistinguishable, and permutations do not generate new
physical microstates.
When this indistinguishability is correctly taken into account, the spurious entropy of
mixing disappears~\cite{swendsen2012introduction}.
In the standard quantum treatment, the paradox is resolved by properly counting
microstates through symmetrization or anti-symmetrization of the wave function.
The total wavefunction of $N$ identical particles satisfies
\cite{zurek2014quantumorigin}
\[
\Psi_{SE}(\mathbf{r}_1, \ldots, \mathbf{r}_N)
= g^{\pi} \psi_{SE}(\mathbf{r}_{\pi(1)}, \ldots, \mathbf{r}_{\pi(N)}),
\tag{27}
\]
where $g^\pi = +1$ for bosons and $-1$ for fermions.
This indistinguishability modifies the canonical partition function as
\[
Z_N = \frac{Z_1^N}{N!}.
\tag{28}
\]

Using the thermodynamic relation
\[
S = \frac{\partial}{\partial T}(k_B T \ln Z_N),
\tag{29}
\]
one obtains the Sackur--Tetrode entropy
\[
S = N k_B \left[ \ln \left( \frac{V}{N \lambda_{\text{th}}^3} \right) + \frac{5}{2} \right],
\tag{30}
\label{42}
\]
which is extensive and resolves the Gibbs paradox.

\subsection{Resolution of the Gibbs Paradox via Quantum Envariance}

From the perspective of quantum envariance, the indistinguishability of identical
particles naturally arises from entanglement between the system and an external
environment that records particle permutations.This perspective does not alter the standard resolution of the Gibbs paradox, but
clarifies its microscopic, information-theoretic origin within quantum mechanics.
The total system--environment state can be represented as a purified,
permutation-symmetric superposition:
\begin{equation}
|\Psi_{SE}\rangle = \frac{1}{\sqrt{N!}} 
   \sum_{\pi} g^{\pi} \,
   |\psi_{SE_{\pi(1)}}\rangle \otimes \cdots \otimes 
   |\psi_{SE_{\pi(N)}}\rangle \otimes |E_{\pi}\rangle,
\tag{31}
\end{equation}
where $|\psi_{SE_{\pi(i)}}\rangle$ denotes the $i$-th single-particle state after
permutation $\pi$, $|E_{\pi}\rangle$ are orthonormal environment states corresponding to
each permutation, and $g^{\pi}=+1$ for bosons or $-1$ for fermions.
This construction ensures invariance of the composite state under particle exchange.

Tracing out the environment yields the reduced density matrix of the physical system:
\begin{equation}
\rho_{N} = \frac{1}{N!}\sum_{\pi} |\Psi_{SE}\rangle\langle\Psi_{SE}|,
\tag{32}
\end{equation}
which represents a maximally mixed state over all $N!$ particle permutations.
The von Neumann entropy of this mixed state is therefore
\begin{equation}
S_{\mathrm{ent}} = -k_{B}\,\mathrm{Tr}(\rho_{N}\ln\rho_{N}) = k_{B}\ln N!.
\tag{33}
\end{equation}
Using Stirling’s approximation, $\ln N! \approx N\ln N - N$, one obtains
\begin{equation}
S_{\mathrm{ent}} \approx N k_{B}(\ln N - 1).
\tag{34}\label{34}
\end{equation}

The term in Eq.~\eqref{34} precisely cancels the \emph{non-extensive entropy of mixing}
that appears in the classical Gibbs paradox.
In the classical treatment, when two identical gases are mixed, the entropy increases by
\[
\Delta S_{\text{mix}} = 2N k_B \ln 2,
\tag{35}
\]
despite the absence of physically distinguishable new microstates.
This spurious contribution arises because classical Maxwell--Boltzmann counting treats
permutations of identical particles as distinct configurations.

In the envariance-based framework, the entropy
$S_{\mathrm{ent}} = k_B \ln N!$
accounts for this permutation redundancy.
Subtracting it from the naive classical entropy restores extensivity:
\[
S_{\text{qm}} = S_{\text{cl}} - S_{\text{ent}}.
\tag{36}
\]
Here $S_{\text{cl}} = N k_B [ \ln ( V / \lambda_{\text{th}}^3 ) + 3/2 ]$ denotes the
entropy obtained under distinguishable-particle counting.
Carrying out the subtraction yields
$
S_{\text{qm}} = N k_B \left[ \ln \left( \frac{V}{N\lambda_{\text{th}}^3} \right)
+ \frac{5}{2} \right],
$
which is identical to the Sackur--Tetrode expression in Eq.~(\ref{42}).

Thus, the entanglement entropy does not introduce a new correction beyond standard
quantum statistical mechanics.
Rather, it provides a microscopic explanation for why the $1/N!$ factor arises when
permutation information is encoded in environmental degrees of freedom and subsequently
traced out.

The key insights are as follows:
\begin{enumerate}
\item The $1/N!$ factor originates from two complementary quantum principles:
(i) permutation symmetry of identical particles, and
(ii) envariance-induced entanglement between the system’s permutation degrees of freedom
and the environment.

\item In the classical (Boltzmann) limit
$(N \lambda_{\text{th}}^3 / V \ll 1)$,
quantum exchange effects vanish and the partition function reduces to
\[
Z_N = \frac{1}{N!} Z_1^N.
\tag{37}
\]

\item The Sackur--Tetrode equation already incorporates this correction implicitly;
the entanglement entropy $S_{\text{ent}}$ should therefore be interpreted as a
\emph{quantum-information--theoretic origin} of the same term.
\end{enumerate}

Thus, the Gibbs paradox is resolved through the combined principles of quantum
indistinguishability and envariance-induced entanglement
\cite{zurek2003,deffner2016,sebastian2019book}.

subsubsection{Validity of Quantum Corrections in Entropy Calculations}
The classical Sackur–Tetrode expression~\cite{sackur1911,tetrode1912} for the entropy of an
ideal monatomic gas does \textit{not} assume that the particles are distinguishable; rather,
it already incorporates the indistinguishability correction through the inclusion of the
$1/N!$ factor in the partition function. This ensures proper counting of microstates and
resolves the Gibbs paradox within the classical framework. However, the Sackur–Tetrode
formula assumes that quantum statistical effects such as exchange correlations and
wavefunction overlap are negligible. This approximation is valid when the thermal
de~Broglie wavelength $\lambda_{\text{th}} = h / \sqrt{2 \pi m k_B T}$ is much smaller than
the mean interparticle spacing. More precisely, quantum corrections are negligible when
the phase-space density satisfies $N \lambda_{\text{th}}^3 / V \ll 1$. Under this dilute,
high-temperature condition, the effects of quantum indistinguishability become
exponentially suppressed, and Eq.~\eqref{42} smoothly reduces to the classical
Sackur–Tetrode form.
\subsubsection{Breakdown of the Classical Approximation at Low Temperatures}

At sufficiently low temperatures or high densities, quantum statistics dominate
\cite{huang1987,pethick2008book,Pitaevskii2016BEC}.
Here ``classical'' refers to the extensive Sackur--Tetrode limit, where the entropy satisfies 
$S(\lambda N,\lambda V,\lambda E)=\lambda S(N,V,E)$. This regime corresponds to 
Maxwell--Boltzmann statistics ($\langle n_k \rangle \ll 1$), which is valid only when quantum exchange correlations are negligible.

Near Bose--Einstein condensation the chemical potential approaches $\mu \to 0^{-}$ and exchange correlations become macroscopic. 
The resulting entropy deviations signal breakdown of the Maxwell--Boltzmann approximation rather than of thermodynamic entropy itself.

\subsubsection{When Are Quantum Corrections Necessary?}

The importance of quantum corrections in thermodynamic quantities can be quantified by
the dimensionless phase-space density, $N \lambda_{\text{th}}^3/V$, where $N$ is the number
of particles, $V$ is the volume, and $T$ is the temperature. This ratio determines whether
the system behaves classically or quantum mechanically.

When $N \lambda_{\text{th}}^3 / V \ll 1$, the wave packets of individual particles scarcely
overlap, and the classical Maxwell–Boltzmann statistics provide an excellent approximation.
In contrast, when $N \lambda_{\text{th}}^3 / V \ge 1$, particle wavefunctions begin to
overlap significantly, and the effects of quantum indistinguishability and exchange
symmetry become non-negligible. In this regime, the Bose–Einstein or Fermi–Dirac statistics
must replace the classical description~\cite{huang1987}. For example, in a typical
room-temperature gas (such as air at $T = 300~\mathrm{K}$ and density
$n \approx 2.5 \times 10^{25}~\mathrm{m^{-3}}$), the ratio $N\lambda_{\text{th}}^3/V$ is
approximately $10^{-7}$, confirming that quantum corrections are completely negligible and
the classical limit holds. However, for an ultracold atomic gas near microkelvin
temperatures ($T \sim 1~\mu\mathrm{K}$), the same parameter can approach values on the order
of $0.3$, signaling the onset of quantum degeneracy where coherence and entanglement effects
become dominant.

Therefore, the quantum correction term—manifested as the $1/N!$ factor in the partition
function or equivalently as the entanglement-induced indistinguishability entropy—becomes
significant only when $\lambda_{\text{th}}$ is comparable to or larger than the mean
interparticle spacing. This condition marks the transition from a classical ensemble
description to a quantum statistical regime~\cite{huang1987,pathria2011}.

\begin{center}
\begin{tabular}{|l|c|l|}
\hline
\textbf{System} & $N\lambda_{th}^3/V$ & \textbf{Effects} \\
\hline
Room-temp gases & $\ll 1$ & Negligible \\
\hline
Cold atoms & $\gtrsim 1$ & Significant \\
\hline
\end{tabular}
\end{center}

The envariance-based symmetry argument has since been extended through
information-theoretic formulations and the framework of quantum resource theories
\cite{jaynes1957information, brukner1999information, cabello2012foundations,
ferrari2025, minagawa2025, d_alessio2015, picatoste2025, popovic2023}.
\subsection{Reformulation of Indistinguishability}

Before concluding this section, let us summarize the structural insight obtained by applying the envariance framework to the Gibbs paradox, in close analogy with the foundational perspective of Deffner and Zurek \cite{deffner2016}. 

Classical treatments of the Gibbs paradox often relied on notions such as ``observer ignorance'' regarding particle identity. Historically, the $1/N!$ factor was introduced to restore extensivity of the entropy, without an explicit microscopic symmetry-based justification. Even in standard quantum treatments, indistinguishability is frequently presented as a counting rule rather than as a structural consequence of symmetry.

Within the envariance framework no appeal to subjective ignorance is required. Particle permutations correspond to transformations that can be compensated by unitary operations on the environment, leaving the global state invariant.

\paragraph{\textbf{Reformulated Fundamental Statement (Gibbs Paradox).}}

\textit{Permutations of identical particles are envariant and therefore physically indistinguishable. The $1/N!$ factor simply reflects that these permutations do not increase the number of accessible states.}

Permutations of identical particles therefore do not enlarge the accessible state space; they represent degenerate symmetry-related descriptions of a single physical configuration. The Gibbs paradox is structurally resolved because quantum symmetry prevents overcounting of states belonging to the same envariant equivalence class.

\section{Entanglement-Based Derivation for Bose--Einstein and Fermi--Dirac Statistics}
\label{Sec.-6}

We now extend the envariance-based~\cite{zurek2003,deffner2016} framework to the
grand-canonical setting, where the system $S$ exchanges both energy and particles with
a large environment acting as a heat and particle reservoir, characterized by temperature
$T$ and chemical potential $\mu$. 

The total system--environment state is written as
\begin{equation}
|\psi_{SE}\rangle
= \sum_{\{n_k\}} C_{\{n_k\}}
\, |\{n_k\}\rangle \otimes |E_{\{n_k\}}\rangle ,
\tag{38}
\end{equation}
where $n_k$ denotes the occupation number of mode $k$, and the environment records the
corresponding configuration.

\paragraph{ Microcanonical global state and envariance.}
The composite system $S+B$ is assumed to be isolated, with fixed total energy
$E_{\mathrm{tot}}$ and total particle number $N_{\mathrm{tot}}$.
According to the principle of envariance~\cite{zurek2003,deffner2016}, all fine-grained
global states within this energy--particle shell are equally weighted, since any swap of
system states can be compensated by a unitary acting on the environment.
This replaces the microcanonical equal--a--priori--probability postulate with a structural symmetry
principle rooted strictly in quantum entanglement.

\paragraph{ Subsystem probabilities from bath degeneracy.}
If the subsystem is found in configuration $\{n_k\}$ with energy
$E_S=\sum_k n_k\epsilon_k$ and particle number $N_S=\sum_k n_k$, the number of compatible
bath states is $\Omega_B(E_{\mathrm{tot}}-E_S,N_{\mathrm{tot}}-N_S)$.
\textbf{Envariance dictates} that subsystem configurations are distinguished only by this bath
degeneracy, as any two configurations with the same $\Omega_B$ must be related by a global unitary swap. Consequently, the statistical weights are fixed as:
\begin{equation}
P(\{n_k\}) \propto \Omega_B(E_{\mathrm{tot}}-E_S,N_{\mathrm{tot}}-N_S) .
\tag{39}
\end{equation}

\paragraph{ Expansion of bath entropy.}
For a macroscopic bath, $\Omega_B = e^{S_B/k_B}$. Expanding to first order around the total system parameters yields
\begin{equation}
S_B(E_{\mathrm{tot}}-E_S,N_{\mathrm{tot}}-N_S)
\simeq S_B(E_{\mathrm{tot}},N_{\mathrm{tot}})
- \beta E_S + \beta \mu N_S ,
\tag{40}
\end{equation}
with
\begin{equation}
\beta = \left(\frac{\partial S_B}{\partial E_B}\right)_{N_B,V_B},
\qquad
\beta \mu = -\left(\frac{\partial S_B}{\partial N_B}\right)_{E_B,V_B} .
\tag{41}
\end{equation}

\paragraph{ Emergence of exponential weighting.}
In this framework, the exponential weights are not postulated but emerge as the objective equilibrium weights compatible with the global symmetry. Substitution of the entropy expansion into the envariance-dictated probability (Eq. 39) yields:
\begin{equation}
P(\{n_k\}) \propto
e^{-\beta(E_S-\mu N_S)}
= e^{-\beta \sum_k n_k(\epsilon_k-\mu)} .
\tag{42--43}\label{42--43}
\end{equation}
Applying the Born rule, the envariance of the global state ensures the reduced density matrix coefficients are:
\begin{equation}
C_{\{n_k\}} \propto
\exp\!\left[
-\frac{\beta}{2}\sum_k n_k(\epsilon_k-\mu)
\right] .
\tag{44}
\end{equation}

\paragraph{ Robustness and Environment Size.}
The derivation of equiprobability via envariance relies on the environment's capacity to perfectly distinguish system states through orthogonal environment states $|\varepsilon_i\rangle$. In macroscopic systems, this condition is naturally satisfied: as the environment's Hilbert-space dimension $D_E$ grows, the overlap between environment states corresponding to different system configurations vanishes as $1/\sqrt{D_E}$. Consequently, the envariance-based equilibrium weighting remains robust even for non-maximal entanglement, provided the environment is sufficiently large to act as a decohering bath.

\paragraph{ Clarification of logical roles: Envariance vs. Exchange Symmetry.}
At this stage, we must emphasize a conceptual separation central to addressing the 
distinction between equilibrium weights and quantum statistics. 
Envariance alone does not determine which subsystem configurations are physically admissible. 
The permissible occupation numbers ($n_k$) of identical particles follow exclusively from 
standard exchange symmetry. The role of envariance in the present framework is strictly to 
fix the objective equilibrium weighting of configurations \emph{within} that symmetry-allowed state space.

\paragraph{ Occupation-number constraints from exchange symmetry.}
Let $\hat{U}_{\pi}^{(S)}$ denote the unitary operator implementing a permutation
$\pi \in \mathfrak{S}_N$ on the subsystem $S$, and $\hat{U}_{\pi}^{(B)}$ the corresponding
compensating transformation acting on the environment $B$.
Envariance under particle exchange requires
$
\hat{U}_{\pi}^{(S)} \otimes \hat{U}_{\pi}^{(B)} \, |\Psi_{SB}\rangle
= |\Psi_{SB}\rangle .
$
Tracing over the environment implies that the reduced equilibrium state is invariant under the action of the permutation group $\mathfrak{S}_N$.

As a consequence, the physical Hilbert space of $N$ identical particles decomposes into
subspaces corresponding to irreducible representations of $\mathfrak{S}_N$.
The symmetric sector permits unrestricted occupation of a single-particle mode,
$
n_k = 0,1,2,\ldots \quad \text{(bosons)},
$
while antisymmetry under exchange enforces the Pauli exclusion principle,
$
n_k \in \{0,1\} \quad \text{(fermions)} .
$
A detailed derivation of these occupation-number constraints based solely on permutation
symmetry, independent of envariance, is provided in Appendix~\ref{AppC}.

\paragraph{ Mode partition function.}
Using Eqs.~\eqref{42--43}, the partition function of a single mode $k$ is
$Z_k = \sum_{n_k} e^{-\beta n_k (\epsilon_k - \mu)}$. Evaluating this sum over the symmetry-allowed values of $n_k$ yields
\begin{equation}
Z_k =
\begin{cases}
\bigl(1 - e^{-\beta(\epsilon_k - \mu)}\bigr)^{-1}, & \text{bosons}, \\[6pt]
\bigl(1 + e^{-\beta(\epsilon_k - \mu)}\bigr), & \text{fermions}.
\end{cases}
\tag{45}
\end{equation}

\paragraph{Average occupation number.}
The mean occupation of mode $k$ is given by
\begin{equation}
\langle n_k \rangle = \frac{1}{Z_k} \sum_{n_k} n_k \, e^{-\beta n_k (\epsilon_k - \mu)} ,
\tag{46}
\end{equation},
which yields 
\begin{equation}
\langle n_k \rangle =
\begin{cases}
\dfrac{1}{e^{\beta(\epsilon_k - \mu)} - 1}, & \text{Bose--Einstein}, \\[8pt]
\dfrac{1}{e^{\beta(\epsilon_k - \mu)} + 1}, & \text{Fermi--Dirac}.
\end{cases}
\tag{47}
\end{equation}

\subsection{Reformulated Fundamental Statement}
Before concluding, let us highlight what has been achieved structurally. Standard approaches to the grand canonical ensemble rely on mathematically ambiguous probabilistic notions such as statistical ensembles, equal a priori probability, or observer ignorance. In contrast, the present approach identifies equilibrium strictly with a symmetry property of entanglement. Following the perspective of Deffner and Zurek\cite{deffner2016,zurek2003decoherence,zurek2009,popescu2006}, we can reformulate the fundamental statement of statistical mechanics for open quantum systems:

\begin{quote}
\textit{
The grand canonical equilibrium of a subsystem $S$ is the reduced state of a globally energy- and particle-degenerate quantum state envariant under all unitaries.
}
\end{quote}

In this framework, the grand canonical equilibrium is not represented by a unique state, but rather by an \textbf{equivalence class} of all maximally envariant states sharing the same total energy and particle number. The resulting Bose--Einstein and Fermi--Dirac distributions do not originate from an independent probabilistic postulate, but emerge naturally as the objective reduced description compatible with this structural symmetry.

\section{Saha Equation with Entanglement-Induced Indistinguishability}\label{Sec.-7}

We revisit the Saha equilibrium for the reaction~\cite{saha1921(2),blundell2009concepts}
\begin{equation}
\mathrm{H} \rightleftharpoons p^{+} + e^{-},
\tag{48}
\end{equation}
and explicitly derive the modified relation incorporating the quantum entanglement correction that enforces indistinguishability through system–environment correlations.

\subsection{Standard Saha equation without quantum correction}

For nondegenerate ideal species, the chemical potential follows from the canonical partition function of an indistinguishable gas,
\begin{equation}
\mu_i = k_B T \ln\!\Big(\frac{n_i\lambda_i^3}{g_i}\Big),
\qquad
\lambda_i = \sqrt{\frac{2\pi\hbar^2}{m_i k_B T}},
\tag{49}
\end{equation}
where $n_i$, $g_i$, and $\lambda_i$ denote number density, internal degeneracy, and thermal wavelength, respectively. Chemical equilibrium, $\mu_H=\mu_p+\mu_e$, yields
\begin{equation}
\frac{n_p n_e}{n_H}
=
\frac{g_p g_e}{g_H}\,\frac{\lambda_H^3}{\lambda_p^3\lambda_e^3}
\,e^{-E_I/(k_B T)}
\tag{50}
\end{equation}
with $E_I$ the ionization energy. Approximating $m_H\simeq m_p\gg m_e$ gives the familiar Saha form~\cite{saha1921(2),blundell2009concepts}
\begin{equation}
\frac{n_p n_e}{n_H}
=
\Big(\frac{2\pi m_e k_B T}{h^2}\Big)^{3/2}
\frac{g_p g_e}{g_H}
\,e^{-E_I/(k_B T)}\tag{51}
\label{eq:saha_standard}
\end{equation}
\subsection{Quantum-entanglement correction}

In a fully quantum description, electrons and protons remain \emph{distinguishable} by intrinsic quantum numbers such as charge and mass. 
However, within each species, individual particles (all electrons or all protons) are \emph{indistinguishable} due to envariance under their mutual permutations. 
The environment effectively records only the exchange information among identical particles of the same type. 
Thus, permutation-equivalent configurations of electrons (or protons) correspond to orthogonal environment states $|E_\pi\rangle$, 
ensuring proper symmetrization within each species without conflating physically distinct particle types. The total system--environment state can then be written as
\[
|\Psi_{SE}\rangle
= \frac{1}{\sqrt{N_e!N_p!}}\sum_\pi
|H,p,e\rangle\otimes|E_\pi\rangle,
\tag{52}
\]
where the sum runs over all particle permutations within each species. 
Tracing out the environmental degrees of freedom removes permutation information, 
so that the number of accessible system microstates is reduced by
\begin{equation}
\Gamma^{\mathrm{ent}}_{\mathrm{ind}} = \frac{1}{N_e! N_p!},
\quad \text{where } N_i = n_i V
\tag{53}\label{53}
\end{equation}
This factor represents an \emph{entanglement-induced indistinguishability correction}:
the environment enforces permutation symmetry, identifying previously overcounted microstates. The associated entanglement entropy is 
\[
S_{\text{ent}} = -k_B\ln \Gamma^{\rm ent}_{\text{ind}}
= k_B\ln(N_e!N_p!),\tag{54}\label{54}
\]
which quantifies the information loss due to tracing over permutation records.  
Since $S_{\text{ent}}\propto \ln N!$, it is weakly (sublinearly) dependent on system size and hence non-extensive, 
consistent with its interpretation as an informational correction rather than a thermodynamic entropy. The corresponding free-energy correction arises from the standard relation $\Delta F = -T\,\Delta S_{\text{ent}}$. 
However, because $S_{\text{ent}}$ scales as $\ln N$, the change $\Delta S_{\text{ent}}$ per particle becomes negligible in the thermodynamic limit, ensuring that the total free energy remains extensive. Thus, $\Delta F$ should be interpreted as a small additive offset that corrects the classical free energy for quantum indistinguishability, not as an extensive contribution.

In summary, the envariance-induced factor $\Gamma^{\rm ent}_{\text{ind}}$ introduces a logarithmic (non-extensive) entropy correction that removes the classical overcounting of identical configurations while preserving the extensivity of macroscopic thermodynamic quantities such as total free energy and entropy.

\subsection{Modified Saha Relation}

\paragraph{Entanglement-induced indistinguishability correction.}
In the envariance-based description, indistinguishability among identical particles arises dynamically through entanglement with the environment. Tracing out the environmental degrees of freedom that record particle permutations reduces the effective number of microstates by
Eq.~\eqref{53}. This modifies the entropy by 
\(S_{\text{ent}} = -k_B\ln\Gamma^{\rm ent}_{\text{ind}}\),
leading to a free-energy shift
\[
\Delta F = -T\,\Delta S_{\text{ent}}
= -k_B T \ln\Gamma^{\rm ent}_{\text{ind}}.
\tag{55}\]
Hence, the equilibrium constant is renormalized as
\(
K_{\text{quantum}} = K_{\text{classical}}\Gamma^{\rm ent}_{\text{ind}},
\)
yielding the modified Saha relation:
\begin{equation}
\frac{n_p n_e}{n_H}
=
\Big(\frac{2\pi m_e k_B T}{h^2}\Big)^{3/2}
\frac{g_p g_e}{g_H}\,
e^{-E_I/(k_B T)}\,
\Gamma^{\rm ent}_{\text{ind}}
\tag{56}\label{56}
\end{equation}
Equation~\eqref{56} reduces to the classical form in the dilute limit 
($\Gamma^{\rm ent}_{\text{ind}}\!\to\!1$), 
while at higher densities 
($\Gamma^{\rm ent}_{\text{ind}}\!<\!1$) 
it predicts a suppression of ionization due to the entanglement-enforced indistinguishability 
of identical particles within each species. 
This expression therefore represents a first-principles, quantum-information-based generalization of the Saha equilibrium, 
with potential implications for dense astrophysical plasmas and recombination-era cosmology.~\cite{milne1924saha, chandrasekhar1958stellar,peebles1993cosmology}.

\subsection{Physical interpretation}

\begin{itemize}
\item \textbf{Classical regime:} when $N_i\lambda_{\text{th}}^3/V\ll1$, particle wave packets do not overlap and $\Gamma_{\text{ind}}\!\approx\!1$.  Eq.~\eqref{eq:saha_standard} is recovered.
\item \textbf{Quantum regime:} for dense or degenerate plasmas ($N_i\lambda_{\text{th}}^3/V\gtrsim1$), environmental entanglement enforces indistinguishability, $\Gamma_{\text{ind}}\!\ll\!1$, and the ionization fraction is suppressed.
\end{itemize}

\section{Results}

This study provides a quantum-information--theoretic perspective on several foundational
results of statistical mechanics within the framework of quantum mechanics, using the
principle of envariance to unify their conceptual origin.

The main results are summarized below:

\begin{enumerate}

    \item \textbf{Derivation of Core Statistical Distributions:}
    By applying the envariance framework~\cite{zurek2003,deffner2016} to a system of $N$
    entangled qubits~\cite{sebastian2019book}, we derived the binomial distribution
    $P(n)=\binom{N}{n}p^n(1-p)^{N-n}$
    from the symmetry of system--environment correlations in maximally envariant equilibrium states.
    The special case of maximal entanglement ($p=1/2$) yields equal probabilities for all
    outcomes, while the general non-maximal case (represented through an auxiliary
    Hilbert space) reproduces Born’s rule~\cite{Born1926ZPhys,zurek2003}.
    In the appropriate limits ($N\!\to\!\infty$, $p\!\to\!0$), this result continuously
    transitions into the Poisson
    $P(n)=\lambda^n e^{-\lambda}/n!$
    and Gaussian distributions, providing a unified quantum-mechanical origin for
    classical statistical laws.

    \item \textbf{Interpretation of the Gibbs Paradox:}
    The Gibbs paradox~\cite{gibbs1902,sackur1911,tetrode1912} was revisited by invoking
    the quantum indistinguishability~\cite{zurek2014quantumorigin} of identical particles.
    We showed that the classical entropy of mixing is exactly cancelled by a von Neumann
    entropy term,
    $S_{\text{ent}} = k_B \ln(N!)$,
    which arises from tracing out environmental degrees of freedom that encode particle
    permutations.
    This entanglement entropy provides a quantum-information--theoretic interpretation
    of the standard $1/N!$ correction already present in the Sackur--Tetrode formula,
    rather than introducing a new modification.
    The relevance of this correction was quantified by the dimensionless parameter
    $N\lambda_{th}^3/V$~\cite{pathria2011,huang1987}.
    Our analysis confirms that this parameter is negligible for a room-temperature gas
    ($\sim 10^{-7}$) but becomes significant for ultracold atomic gases
    ($\sim 0.3$), thereby delineating the classical-to-quantum crossover.

    \item \textbf{Emergence of Bose--Einstein and Fermi--Dirac Distributions:}
    Extending the envariance principle to the grand canonical ensemble shows that
    equilibrium quantum statistics arise from the combination of
   exchange symmetry (assumed as an input from quantum mechanics)
    and envariance-based weighting of configurations.
    For a composite state
    $|\psi_{SE}\rangle = \sum_{\{n_k\}} C_{\{n_k\}}
    |\{n_k\}\rangle\otimes |E_{\{n_k\}}\rangle$,
    invariance under local unitaries acting on $S$ fixes the amplitudes as
    $C_{\{n_k\}} \propto \exp[-\tfrac{\beta}{2}\sum_k n_k(\epsilon_k-\mu)]$.
    This leads to the probability distribution
    $P(\{n_k\})\propto e^{-\beta\sum_k n_k(\epsilon_k-\mu)}$
    and hence to the Bose--Einstein and Fermi--Dirac occupation numbers
    $\langle n_k\rangle = 1/[e^{\beta(\epsilon_k-\mu)} \mp 1]$
    \cite{nikhil2021statistical,sur2023quantum}.
    Envariance thus fixes the equilibrium weighting of symmetry-allowed configurations,
    while the exchange symmetry itself determines the allowed occupation numbers.

    \item \textbf{Entanglement-Based Interpretation of the Saha Equation:}
    Incorporating indistinguishability within each particle species through
    environment-induced entanglement yields an
    effective quantum-information--based reformulation
    of the Saha equilibrium for ionization~\cite{saha1921(2),mihalas1978},
    as shown in Eq.~\eqref{54},
    where $\Gamma_{\text{ind}} = 1/(N_e!\,N_p!)$ is an entanglement-induced
    indistinguishability factor.
    This factor represents the same permutation correction already implicit in
    quantum statistical mechanics and leads to a suppression of the effective
    ionization fraction at high densities.
    In the dilute (classical) limit $\Gamma_{\text{ind}}\to1$, the conventional Saha
    equation is recovered, while in dense or partially degenerate regimes the correction
    becomes non-negligible,
    consistent with known exchange and quantum-degeneracy effects in plasmas.

    This overall picture is consistent with modern resource-theoretic and typicality
    approaches~\cite{lostaglio2019coherence,yunger2020typicality,
    faist2021thermodynamicreversibility,weilenmann2023quantumthermo,
    wilming2019entanglemententropy},
    which regard entropy, equilibration, and thermal behavior as consequences of
    information-theoretic constraints.
\end{enumerate}

\section{Discussion}

The results presented in this paper offer a significant shift in perspective, moving the
foundations of statistical mechanics from a postulate-based framework
toward a formulation in which its core structures admit a deductive
interpretation within quantum mechanics.
The central implication is that statistical laws are not merely abstract rules for dealing
with ignorance but are physical consequences of the way quantum
information~\cite{sebastian2019book} is structured and processed through entanglement
\cite{deffner2016,zurek2003} with an environment.

Our derivation of the core ensembles~\cite{pathria2011} via envariance reframes the role of
the observer and the environment.
Rather than invoking probabilities as external assumptions, the
environment actively enforces the statistical structure of equilibrium states.
The equiprobability of the microcanonical ensemble arises from the maximal symmetry
\cite{zurek2014quantumorigin} of a system--environment state, while the exponential form of the canonical
ensemble~\cite{deffner2016} emerges as the statistically inevitable outcome for any small
part of that larger whole.
This provides a physical mechanism for thermalization and equilibrium, rooted in the flow
and redistribution of quantum information~\cite{sebastian2019book}.

The unified derivation of the binomial, Poisson, and Gaussian distributions from a single
quantum-mechanical model underscores the predictive power and internal coherence of this
methodology.
It shows that the statistics that govern discrete events (like particle counts) and
continuous variables are not separate things, but arise from common underlying
combinatorial and entanglement structures.
From this perspective, classical probability theory appears as an
effective description emerging from more fundamental quantum-information constraints
\cite{zurek2003}.

Furthermore, the resolution of the Gibbs paradox~\cite{gibbs1902,sackur1911,tetrode1912}
through entanglement~\cite{deffner2016} entropy provides more than just a fix for a classical
problem.
It offers a quantitative tool, the parameter $N\lambda_{\text{th}}^3/V$,
\cite{pathria2011,huang1987} to delineate the boundary between the classical and quantum
worlds.
Our numerical examples for air versus ultracold atomic gas make this transition tangible,
showing that the ``paradox'' is simply the breakdown of a classical approximation in a
regime where quantum effects, specifically entanglement due to indistinguishability
\cite{zurek2014quantumorigin}, cannot be ignored.

It is essential to acknowledge the idealizations in our models, such as the assumptions of
non-interacting particles in the derivation of the Gibbs paradox
\cite{gibbs1902,sackur1911,tetrode1912} and the perfectly orthogonal environment states.
Future research works could investigate the applications of this envariance-based
framework~\cite{deffner2016} in strongly interacting systems, where the correlations may
induce new statistical phenomena and phase transitions~\cite{huang1987,pathria2011}.
Analyzing the effects of imperfect environmental coupling and decoherence would produce a
more accurate depiction of the statistical equilibrium in open quantum systems.
Finally, the success of this approach suggests its potential relevance to other enduring
issues at the intersection of information theory, gravitation, and thermodynamics, such as
the black hole information paradox.

This work establishes that the central principles of statistical mechanics
admit a consistent interpretation within quantum foundations, without introducing new
phenomenological assumptions beyond those already present in quantum mechanics.
By employing the concept of envariance~\cite{deffner2016,zurek2003}—a symmetry property of
entangled quantum systems—we have shown how the microcanonical and
canonical ensembles~\cite{pathria2011}, along with their characteristic probability
distributions, emerge naturally from the structure of quantum information
dynamics rather than being imposed as independent axioms.

The key insight unifying our results is that thermodynamic behavior~\cite{sebastian2019book}
arises from the information-theoretic properties of quantum systems interacting with their
environment.
This perspective provides a natural resolution to the Gibbs paradox
\cite{gibbs1902,sackur1911,tetrode1912} through entanglement entropy, while simultaneously
explaining the emergence of classical statistical distributions from quantum measurement
processes.
The quantum corrections we derive for the Saha equation~\cite{saha1921(2),mihalas1978}
further demonstrate how this framework extends beyond textbook statistical mechanics to
predict observable phenomena in quantum many-body systems~\cite{nielsen2000}.
These findings demonstrate that:

\begin{enumerate}
\item Thermodynamic laws are emergent consequences of quantum information structure.
\item Classical paradoxes disappear when proper account is taken of quantum correlations.
\item The boundary between quantum and classical statistical behavior can be precisely
quantified.
\end{enumerate}

This quantum foundations approach opens up new ways to study complex systems where
information, thermodynamics, and quantum dynamics intersect.
The techniques formulated herein may be utilized in domains extending from quantum
thermodynamics to the investigation of emergent phenomena in condensed matter systems.
Rather than replacing existing theory, the present framework provides a
coherent conceptual bridge between mechanics and thermodynamics, complementing the program
initiated by Boltzmann.
Recent work in quantum thermodynamics and decoherence supports this envariance-based
framework~\cite{zurek2009quantum, riedel2010quantum, picatoste2025, popovic2023,
ferrari2025}.

\section{Acknowledgments}
A.G acknowledges partial financial support from IITK. Amul and Shubhit acknowledge partial support from IISER Bhopal.

\appendix
\section{Poisson distribution as limit of the binomial distribution}

Taking the limiting case, Eq.~\eqref{16} can be derived from Eq.~\eqref{15}.
We substitute $p = \lambda/N$ into the binomial formula to obtain:
\begin{equation}
P(n) = \binom{N}{n} \left(\frac{\lambda}{N}\right)^n
\left(1 - \frac{\lambda}{N}\right)^{N-n}.
\tag{A1}
\end{equation}
Next, we expand the binomial coefficient
$\binom{N}{n} = \frac{N!}{n!(N-n)!}
= \frac{N(N-1)\cdots(N-n+1)}{n!}$.
Thus,
\begin{equation}
\begin{split}
P(n) &= \frac{\lambda^n}{n!}
      \left[ \frac{N(N-1)\cdots(N-n+1)}{N^n} \right] \\
     &\quad \times \left(1 - \frac{\lambda}{N}\right)^N
      \left(1 - \frac{\lambda}{N}\right)^{-n}.
\end{split}
\tag{A2}
\end{equation}

Now we take the limit as $N \to \infty$:
\begin{enumerate}
    \item
    $\displaystyle
    \lim_{N \to \infty}
    \frac{N(N-1)\cdots(N-n+1)}{N^n}
    = 1,
    $
    since each factor approaches unity.
    \item
    $\displaystyle
    \lim_{N \to \infty}
    \left(1 - \frac{\lambda}{N}\right)^N
    = e^{-\lambda}.
    $
    \item
    $\displaystyle
    \lim_{N \to \infty}
    \left(1 - \frac{\lambda}{N}\right)^{-n}
    = 1.
    $
\end{enumerate}
Combining these results yields
\begin{equation}
\lim_{N \to \infty} P(n)
= \frac{\lambda^n e^{-\lambda}}{n!}.
\tag{A3}
\end{equation}
This is precisely the Poisson distribution given in Eq.~\eqref{16}
\cite{feller1968}.
As required for a normalized probability distribution,
\[
\sum_{n=0}^{\infty} \frac{\lambda^n e^{-\lambda}}{n!}
= e^{-\lambda} \sum_{n=0}^{\infty} \frac{\lambda^n}{n!}
= 1.
\tag{A4}
\]

\section{Gaussian (de Moivre--Laplace) limit}

Define $\lambda = Np$ and $\sigma^2 = Np(1-p)$.
We approximate $P(n)$ for large $N$ using Stirling's approximation
\[
\ln n! \simeq n\ln n - n + \tfrac{1}{2}\ln(2\pi n)
+ \mathcal{O}(1/n).
\tag{B1}
\]
We write
\[
\ln P(n) = \ln\binom{N}{n}
+ n\ln p + (N-n)\ln(1-p).
\tag{B2}
\]
Using Stirling’s approximation for the factorials and expanding to second order
around $n=\lambda$ (i.e., setting $n=\lambda+\delta$ and assuming $|\delta|\ll N$),
the linear term vanishes and the quadratic term yields
\[
\ln P(n) \approx \text{const}
- \frac{(n-\lambda)^2}{2\sigma^2}.
\tag{B3}
\]
Exponentiating and restoring the normalization factor gives the Gaussian form
\begin{equation}
\begin{split}
P(n) &\approx
\frac{1}{\sqrt{2\pi\sigma^2}}
\exp\!\left[-\frac{(n-\lambda)^2}{2\sigma^2}\right] \\
&=
\frac{1}{\sqrt{2\pi Np(1-p)}}
\exp\!\left[-\frac{(n-Np)^2}{2Np(1-p)}\right].
\end{split}
\tag{B4}
\end{equation}

\section{Derivation of Occupation-Number Constraints from Exchange Symmetry}
\label{AppC}

In this appendix, we explicitly derive the allowed occupation numbers for identical
particles in three spatial dimensions. The purpose is to clarify, step by step,
why symmetric exchange symmetry permits unrestricted occupation numbers
$n_k = 0,1,2,\dots$, while antisymmetric exchange symmetry enforces the Pauli exclusion
principle $n_k \in \{0,1\}$.
This derivation relies solely on standard quantum-mechanical permutation symmetry and
does not invoke envariance itself. Envariance enters in the main text only to fix the
equilibrium weighting of the symmetry-allowed configurations.

\subsection{Permutation symmetry and identical particles}

Consider a system of $N$ identical particles with single-particle Hilbert space
$\mathcal{H}_1$. The $N$-particle Hilbert space is the tensor product
\begin{equation}
\mathcal{H}^{(N)} = \mathcal{H}_1^{\otimes N}.
\end{equation}
For any permutation $\pi \in \mathfrak{S}_N$ of particle labels, let
$\hat{U}_\pi$ denote the corresponding unitary operator acting on
$\mathcal{H}^{(N)}$:
\begin{equation}
\hat{U}_\pi
\big( |\psi_1\rangle \otimes \cdots \otimes |\psi_N\rangle \big)
=
|\psi_{\pi(1)}\rangle \otimes \cdots \otimes |\psi_{\pi(N)}\rangle.
\end{equation}

Identical particles are defined by the requirement that all physical observables
commute with $\hat{U}_\pi$. Consequently, physically admissible states must transform
under irreducible representations of the permutation group
$\mathfrak{S}_N$ \cite{landau1980,mandl1988,weinberg1995}.

\subsection{Symmetric and antisymmetric subspaces}

In three spatial dimensions, consistency with locality and relativistic causality
restricts admissible representations to the one-dimensional irreducible
representations of $\mathfrak{S}_N$:
\begin{align}
\hat{U}_\pi |\Psi\rangle &= + |\Psi\rangle
\quad \text{(symmetric sector, bosons)}, \\
\hat{U}_\pi |\Psi\rangle &= (-1)^{\pi} |\Psi\rangle
\quad \text{(antisymmetric sector, fermions)},
\end{align}
where $(-1)^{\pi}$ denotes the parity of the permutation.
Accordingly, the physical Hilbert space decomposes as
\begin{equation}
\mathcal{H}^{(N)} =
\mathcal{H}_{\mathrm{sym}} \oplus \mathcal{H}_{\mathrm{antisym}}.
\end{equation}

\subsection{Occupation-number basis}

Let $\{ |k\rangle \}$ denote a complete orthonormal basis of single-particle eigenstates.
A general many-body state may be described by occupation numbers
$\{n_k\}$ satisfying
\begin{equation}
\sum_k n_k = N.
\end{equation}
In second-quantized language, the corresponding Fock states are
\begin{equation}
|\{n_k\}\rangle =
\prod_k \frac{(a_k^\dagger)^{n_k}}{\sqrt{n_k!}} |0\rangle,
\end{equation}
where $a_k^\dagger$ creates a particle in mode $k$.

\subsection{Bosons: symmetric sector}

For bosons, the many-body wavefunction is symmetric under all permutations.
Equivalently, bosonic creation operators satisfy the commutation relations
\begin{equation}
[a_k, a_{k'}^\dagger] = \delta_{k k'}.
\end{equation}
Because the operators commute,
\begin{equation}
(a_k^\dagger)^n |0\rangle \neq 0
\quad \forall\, n \in \mathbb{N},
\end{equation}
so that
\begin{equation}
n_k = 0,1,2,\dots \qquad \text{(bosons)}.
\end{equation}

\subsection{Fermions: antisymmetric sector}

For fermions, the wavefunction is antisymmetric under exchange.
The creation and annihilation operators satisfy
\begin{equation}
\{a_k, a_{k'}^\dagger\} = \delta_{k k'}.
\end{equation}
In particular,
\begin{equation}
(a_k^\dagger)^2 = 0,
\end{equation}
which enforces
\begin{equation}
n_k \in \{0,1\} \qquad \text{(fermions)}.
\end{equation}

\subsection{Summary and relation to envariance}

The occupation-number constraints
\begin{equation}
n_k =
\begin{cases}
0,1,2,\dots & \text{bosons}, \\
0,1 & \text{fermions},
\end{cases}
\end{equation}
are direct consequences of permutation symmetry in quantum mechanics
\cite{landau1980,mandl1988}.

Within the envariance-based framework of the main text, these exchange-symmetry sectors
play a complementary role. Exchange symmetry determines the admissible configuration
space, while envariance determines which symmetry-allowed configurations are
equilibrium-stable and assigns them equal statistical weight within the corresponding
sector.

In this sense, envariance does not generate Bose--Einstein or Fermi--Dirac statistics,
but it explains why only fully symmetric or antisymmetric exchange sectors yield
self-consistent equilibrium ensembles for macroscopic systems, and why no additional
combinatorial postulates are required once the symmetry sector is fixed.

\bibliographystyle{apsrev4-2}
\bibliography{ref}

\end{document}